\documentclass[]{aa} 
\usepackage{graphicx}
\usepackage{txfonts}
\usepackage{natbib}
\bibpunct{(}{)}{;}{a}{}{,}
\usepackage{longtable}

\begin{document}

\newcounter{subfigure}

\title{Quiet Sun magnetic fields observed by Hinode: Support for a local dynamo}

\author{D. Buehler \inst{1}$^,$\inst{2}, A. Lagg \inst{1} \and S.K. Solanki \inst{1}$^,$\inst{3}}

\institute{\inst{1} Max Planck Institute for Solar System Research, Max-Planck-Stra\ss e 2, 37191 Katlenburg-Lindau, Germany\\
\inst{2} 
Georg August University, Institute for Astrophysics, Friedrich-Hund-Platz 1,
37077 G\"ottingen, Germany\\
\inst{3} School of Space Research, Kyung Hee University, Yongin, Gyeonggi, 446-701, Korea}

\date{Received January 30, 2013; accepted April 24, 2013}

\abstract
{The Hinode mission has revealed copious amounts of horizontal flux covering the quiet Sun.  Local dynamo action has been proposed to explain the presence of this flux. }
{We sought to test whether the quiet Sun flux detected by Hinode is due to a local or the global dynamo by studying long-term variations in the polarisation signals detectable at the disc centre of the quiet Sun between November 2006 and May 2012, with particular emphasis on weak signals in the internetwork.}
{The investigation focusses on line-integrated  circular polarisation $V_{tot}$ and linear polarisation $LP_{tot}$ profiles obtained from the Fe I 6302.5$\,\AA$ absorption line in Hinode SOT/SP.}
{Both circular and linear polarisation signals show no overall variation in the fraction of selected pixels from 2006 until 2012. There is also no variation in the magnetic flux in this interval of time. The probability density functions (PDF) of the line-of-sight magnetic flux can be fitted with a power law from $1.17 \times 10^{17}$ Mx to $8.53 \times 10^{18}$ Mx with index $\alpha=-1.82\pm0.02$ in 2007. The variation of $\alpha$'s across all years does not exceed a significance of $1\sigma$. Linearly  polarised features are also fitted with a power law, with index $\alpha=-2.60\pm0.06$ in 2007. Indices derived from linear polarisation PDFs of other years also show no significant variation.}
{Our results show that the ubiquitous horizontal polarisation on the edges of bright granules seen by Hinode are invariant during the minimum of cycle 23. This supports the notion that the weak circular and linear polarisation is primarily caused by an independent local dynamo.}


\maketitle

\section{Introduction}

Quiet Sun observations made by the SOT \citep{tsuneta2008sot,suematsu2008sot,ichimoto2008sot,shimizu2008sot} on Hinode \citep{kosugi2007} as well as IMaX \citep{martinez2011} on $\emph{\sc Sunrise}$ \citep{solanki2010,barthol2011} have revealed copious amounts of horizontal magnetic flux. These observations, using both Hinode SOT/SP and SOT/NFI, also showed that this horizontal magnetic flux is concentrated in patches on the edges of bright granules \citep{lites2008,ishikawa2008}. \citet{danilovic2010sunrise} found using $\emph{\sc Sunrise}$ that the lifetime of these flux patches ranges from one minute right up to the granular lifetime of five minutes. The apparent isolation of these flux patches from other quiet Sun phenomena have made them strong candidates for being generated by local dynamo action just below the photosphere \citep{cattaneo1999,danilovic2010,ishikawa2009}. Also, using SOT/SP data from plage regions and quiet Sun, \citet{ishikawa2008} showed that these fields have no preferred orientation and are independent of the strength of nearby magnetic fields. The flux distributions of magnetic features obtained from SOHO/MDI and Hinode SOT/NFI have also been interpreted in terms of a turbulent dynamo occurring continuously over a range of scales throughout the convection zone \citep{parnell2009,thornton2011}.\\

In spite of the evidence presented in the papers above, the true source of the weak magnetic flux features in the quiet Sun is still not resolved \citep{petrovay1993}. In particular it is unclear if they are produced by a turbulent local dynamo \citep{voegler2007,schuessler2008,danilovic2010,pietarila2010}, with predominantly horizontal magnetic fields \citep{lites2008,orozco2007} or due to the global dynamo, e.g. field recycled from decaying active and ephemeral active regions. \citet{asensio2009} concludes that quiet Sun magnetic fields observed by Hinode have an isotropic distribution and \citet{stenflo2010} finds predominantly vertical fields speaking against a local dynamo. \citet{stenflo2012} argues using SOHO/MDI data that the global dymano is the main source of magnetic flux even in the quiet Sun. Magnetic flux could also be generated from rising magnetic loops that \emph{"explode"} within the convection zone \citep{rempel2001}. A combination of different sources is also conceivable. \\
One way to distinguish between global and local dynamos is to consider the variation of the magnetic flux over the solar cycle, since, put simply, flux due to a local dynamo is not expected to vary over the solar cycle, while flux produced by the global dynamo should display a cyclic variation. There is some uncertainty regarding the magnitude of this variation due to the fact that smaller magnetic bipoles, such as ephemeral active regions, show a rather low cycle amplitude, with shifted phase relative to the strong cyclic variation of active regions \citep{harvey1993,hagenaar2001,jin2012}.\\
The Hinode satellite has recorded images of the quiet Sun for over six years now, covering the extraordinary solar minimum of cycle 23. Solar activity during this minimum was markedly lower than in the minima of previous solar cycles \citep{lockwood2010,froehlich2009}. Although this period is considerably shorter than a full solar cycle, we nonetheless use all available Hinode SOT/SP quiet Sun disk centre scans over this period to address the question whether solar cycle variations can be measured down to granular scales.

\section{Data and method of analysis}

Here we focus entirely on the quiet Sun observed close to the solar disc center from November 2006 to May 2012 and Hinode SOT/SP images were chosen accordingly. In this Section we present the data employed and describe the various issues that presented themselves and how they were accounted for. \\

Since this investigation is concerned with the weak polarisation signals in the quiet Sun, a good Signal-to-Noise ratio is paramount. Therefore, images with an exposure time of at least 4.8 s operated in dual mode were preferred. However, due to telemetry problems starting in 2008 dual mode images were only available until  January 2008. Images recorded after January 2008 were single mode images. To increase the statistical significance and to minimise selection effects, images covering a large area of the solar disc were preferred, ranging up to 320'' $\times$ 160''. \\
Apart from 2007 the number of images satisfying these criteria was rather patchy, so that images recorded as part of the irradiance program were used (HOP 79).  Images belonging to HOP 79 were shot almost every month and cover the quiet Sun at the disc center. However, two issues emerged during the investigation while working with these images. Firstly they received an onboard binning in the slit direction, which decreased the resolution and required, for consistency, all the other images used in this investigation to be binned retrospectively. Also, the images in the irradiance program are comparatively small, only 30'' $\times$ 120'' each. This was partly compensated by using two images from the irradiance program recorded on the same day, but on different locations around the disc center, thereby increasing the considered area to 30'' $\times$ 240''. 
Table \ref{SOTimg} (available electronically) gives an overview of all the images used in this investigation. Every image was calibrated using the \emph{sp\_prep} routine of the solarsoft package, followed by a bitshift correction of the four Stokes parameters that mainly affected images recorded in dual mode.

\subsection{Instrumental effects}

\begin{figure*}
   \centering
   \begin{minipage}{0.5\linewidth}
\centering
\includegraphics[width=6.5cm]{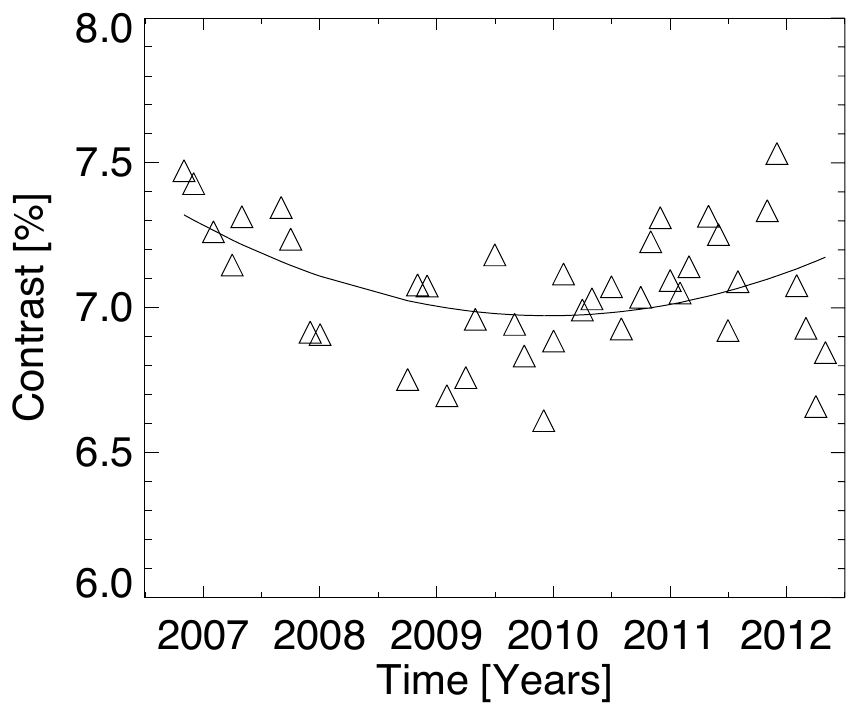}
\end{minipage}%
\begin{minipage}{0.5\linewidth}
\centering
\includegraphics[width=6.5cm]{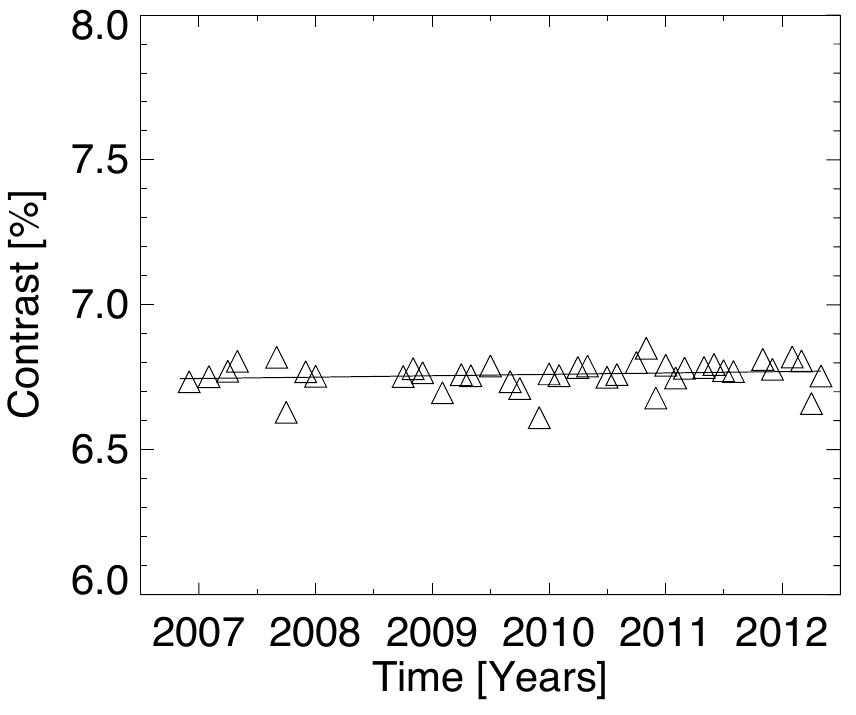}
\end{minipage} 
   \caption{\emph{Left:} Original Stokes $I$ continuum contrast variation of the employed SOT/SP images. The solid line is a quadratic fit with quadratic and linear coefficients of 0.00025 $\pm$ 0.00009 and -0.019 $\pm$ 0.006. \emph{Right:} Stokes $I$ continuum contrast variation after each image was convolved using an artificial defocus. The solid line is a constant drawn at 6.75\%.}
    \label{spcontr}
    \end{figure*}

The analysis of the calibrated images was performed by a pixel selection algorithm. Pixels were selected according to a preset uniform polarisation threshold. Finding a robust threshold proved to be a delicate process, because images recorded in 2007 in dual mode have a smaller noise level by a factor of $\sqrt{2}$ than single mode images recorded in 2008. Images from the irradiance program have a different noise level yet again due to the onboard binning and subsequent \emph{jpeg} compression. This problem was minimised by reducing calibrated dual mode images from 2007 to single mode images and subsequently binning them. After this process was completed every image had the same effective pixel size (0.16'' $\times$ 0.32'') as well as the same, within one significant figure, standard deviation ($\sigma$) noise level for the Stokes $Q$, $U$ and $V$ parameters estimated from the continuum pixels. The standard deviation ($\sigma$) was found to increase by 10\% over six years starting from a value of $\sigma =0.9 \times 10^{-3}$ $I_{c}$ in November 2006. \\
The $\emph{rms}$ contrast of Stokes $I$ was also calculated to discern any variation in the PSF over time. We found that it takes an average value of $7\pm0.3\%$. This is in good agreement with the results of \citet{danilovic2008}, who also also obtained a value of 7\% prior to the deconvolution by the instrument's PSF. However, the continuum contrast, shown in Fig. \ref{spcontr}, is not constant over the period of investigation. It can vary by as much as $\approx$1\% and is caused by temperature fluctuations on the spacecraft (Katsukawa private communication). This variation also affects the Stokes $Q$, $U$ and $V$ profiles in an image, analyzed in Sects. 3 and 4, and therefore needs to be compensated.

Every image was convolved by an appropriate artificial defocus using an MTF generated by the commercial software package $\emph{ZEMAX}$ following \citet{danilovic2008}. The convolution was applied to all 112 wavelength positions of each Stokes parameter. The result of this convolution on the $\emph{rms}$ contrast can also be seen in Fig. \ref{spcontr} where the remaining scatter is the result of the applied artificial defocus being in steps of 0.17mm similar to the defocus steps available on the Hinode spacecraft.\\
The variation of $\emph{rms}$ contrast not only affects SOT/SP but has also been observed with the filtergraph aboard Hinode. \citet{muller2011} found a reduction in the continuum contrast using images from SOT/BFI between November 2006 until July 2010. By using wide band continuum images recorded at 6684$\AA$ we not only found a decrease in the continuum contrast until July 2010, but a subsequent increase until Jan 2012 amounting to a variation of $\approx$1\%. Unlike the spectrograph, the filtergraph has regular focus campaigns to ensure that the instrument is in the optimum focus position. Since both instruments show the same variation within error this rules out the possibility that the contrast variation seen in Fig. \ref{spcontr} is merely the result of a relative shift between the filtergraph and the spectrograph. 
    
   \begin{figure}
   \centering
   \includegraphics[width=7cm]{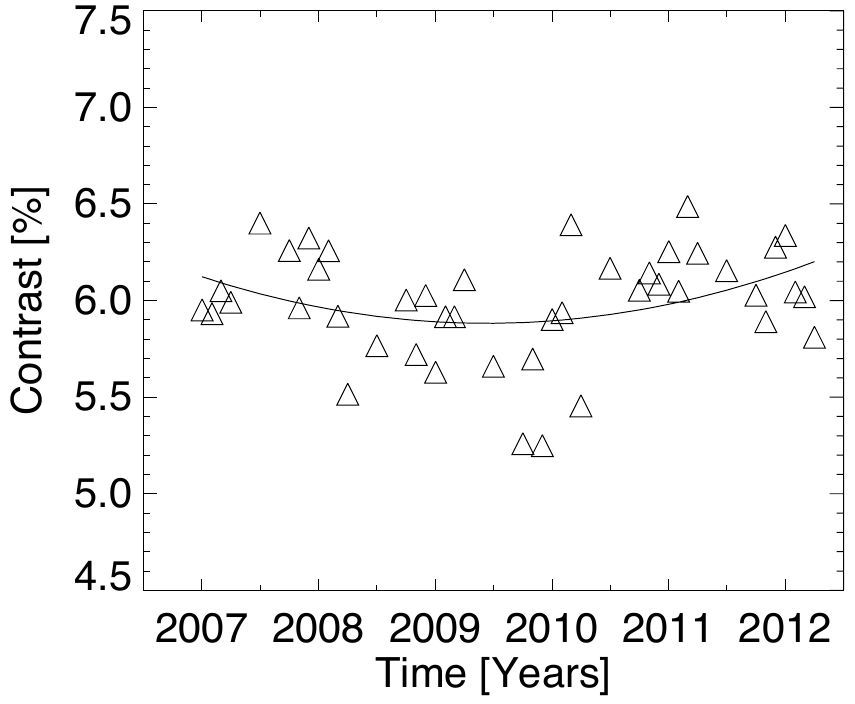}
      \caption{Contrast variation of SOT/BFI red continuum images at 6684$\AA$. The solid line is a quadratic fit with quadratic and linear coefficients of $0.00028 \pm 0.00013$ and $-0.017 \pm 0.009$.}
         \label{sotwbcontr}
   \end{figure}

\subsection{Image analysis}

In order to avoid any biases introduced by inversions see \citep{borrero2011}, we analysed the observed Stokes profiles directly. In parts of this investigation an estimate of the magnetic flux density was made, based on the observed line-of-sight polarisation signals. This was done using the standard magnetograph formula \citep{unno1956}. However, this formula suffers from the assumption of a weak magnetic field, which is not fulfilled by network magnetic elements and not even by all magnetic features in the internetwork \citep{lagg2010}. This formula also assumes certain atmospheric parameters to be constant, such as the line-of-sight velocity and the magnetic field strength, which is not true for weak quiet Sun Stokes profiles \citep{viticchie2010}. Therefore the magnetic flux density value returned by the formula must be considered as an approximation only. In addition, this analysis assumes that the magnetic feature is resolved, so that the flux density is underestimated if this assumption is not fulfilled. The estimate of magnetic flux of the feature should not be affected, however.\\ 
Several steps were taken to improve the Signal-to-Noise ratio and thus increase the number of pixels above a given threshold. The entire investigation focussed only on the Fe I 6302.5$\,\AA$ absorption line as it is a Zeeman triplet and has a larger Land\'e factor ($g = 2.5$) compared to the Fe I 6301.5$\,\AA$ absorption line. It, therefore, is more sensitive to weak magnetic fields. The total circular polarisation signal was calculated using $M$ = 9 wavelength points around the central wavelength of the Fe I 6302.5$\,\AA$ absorption line.

\begin{equation}
V_{tot} = \frac{1}{ M} \sum^M_1 {{|}V{|}}.
\end{equation}

The absolute value of the Stokes $V$ signal was taken at each wavelength point to account for the full polarimetric signal. The fraction of pixels displaying significant circular polarisation, $P_{sel} (V_{tot})$, in an image could then be calculated. The average non-zero $\emph{rms}$ noise of these profiles was removed from $V_{tot}$ before being compared to a threshold. The magnitude of the offset was calculated for each profile individually using wavelength points located in the continuum of the spectrum.\\ 
Instead of using the individual Stokes $Q$ and $U$ profiles, the total linear polarisation, $LP$, was calculated and is defined as, 
\begin{equation}
LP = \frac{\sqrt{Q^2 + U^2}}{I_c}.
\end{equation}
Although increasing the signal-to-noise ratio, this process meant that any information on the azimuth of the magnetic vector contained in the polarisation signal was lost. Finally, the $LP$ profile of the Fe I 6302.5$\,\AA$ line was integrated along the dispersion direction using,
\begin{equation}
LP_{tot} = \frac{1}{ M} \sum^M_1 {LP}.
\end{equation}
The mean non-zero rms noise that arises in the calculation of $LP_{tot}$ was removed for each of the $M$ = 9 wavelength  points before a constant linear polarisation threshold for $LP_{tot}$ was set. Then $P_{sel} (LP_{tot})$, the fraction of pixels selected on the basis of the above criteria, could be determined. \\

   \begin{figure*}
   \centering
   \includegraphics{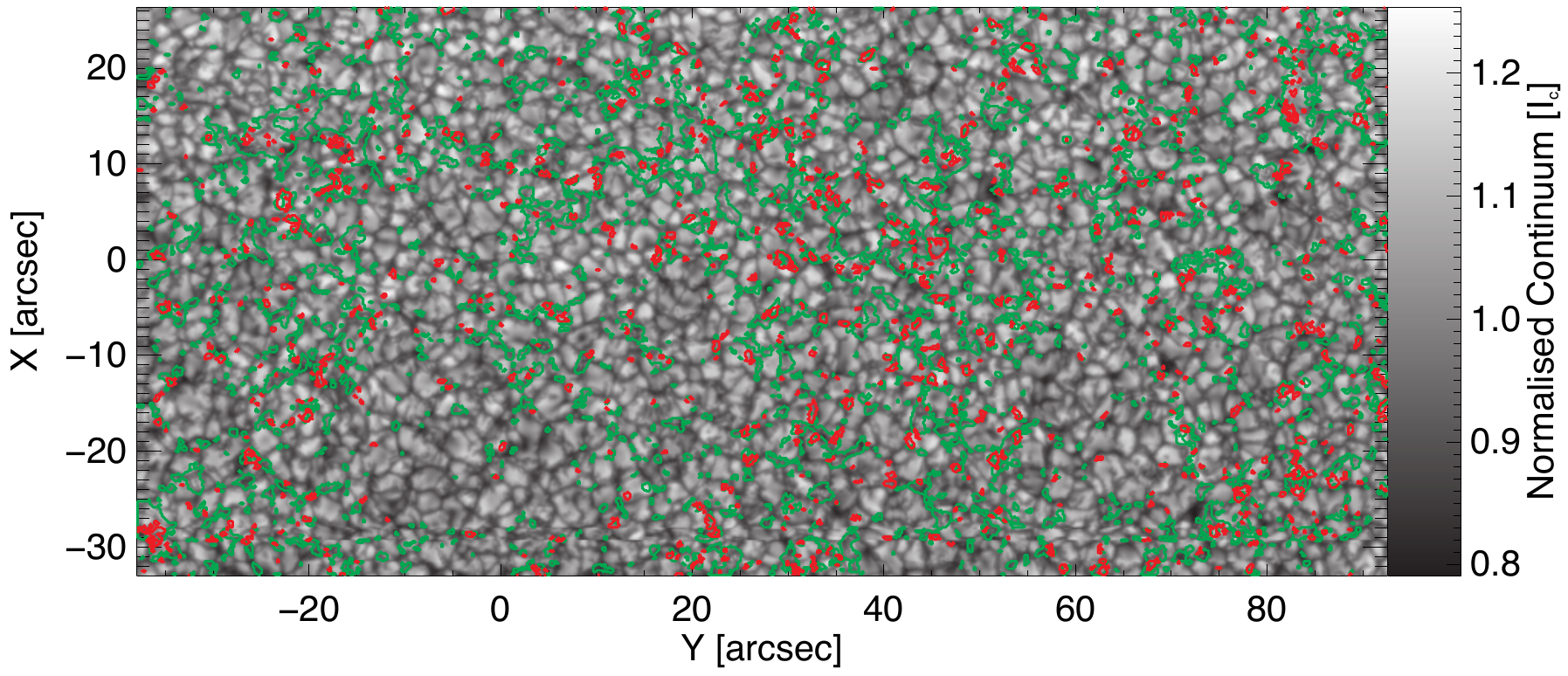}
   \caption{Quiet Sun continuum image from 30$^{th}$ January 2008 UT 23:34:05 with overplotted contours of both circular and linear polarisation patches corresponding to $\approx$$7\sigma$ for circular polarisation (green) and $\approx$$3\sigma$ for linear polarisation (red). The image was subjected to an 2x1 binning in the slit direction. Each contour encompasses a patch of at least two pixels in size.}
    \label{picture}
    \end{figure*}

The fraction of pixels lying above this threshold is called the occurrence of the circular, $P_{sel} (V_{tot})$, and of the linear, $P_{sel} (LP_{tot})$, polarisation.
Patches of linear polarisation were analysed separately from those of circular polarisation. One reason for doing so was due to the different locations on an arbitrary quiet Sun image where patches of different polarisation appeared. Patches of strong circular polarsiation were almost exclusively found in the intergranular lanes, whereas patches of linear polarisation preferred the edges of granules \citep{lites2008,ishikawa2008}. Figure \ref{picture} is a typical quiet Sun image used in this investigation with the contours showing regions with significant polarisation signals (see caption for details).\\

The results presented in Sections 3 \& 4 were obtained by setting a threshold of $V_{tot} = 0.3\% (\approx7\sigma)$ on the circular polarisation and $ LP_{tot} = 0.095\% (\approx3\sigma)$ on the linear polarisation, respectively. The same threshold was chosen for all images. Isolated pixels that met a particular polarisation threshold were discarded, to prevent the selection of pixels caused by noise spikes. This precaution decreased the number of selected pixels in an image by 15\%-20\%, so that on average 11\% of pixels in an image carried significant circular polarisation and 3\% significant linear polarisation. The exact values of the chosen thresholds do not change the results significantly, as we learnt by trying out different values. However, raising the value of the threshold for $LP_{tot}$ led to a rapid decrease in the number of chosen pixels and thus to significantly poorer statistics, while for $V_{tot}$ this is less of an issue. However, small thresholds for $V_{tot}$ compromised the patch size analysis presented in the following sections by merging the many of the strong flux concentrations in an image. Since our conclusions are not affected, descriptions of analyses employing higher or lower thresholds have been omitted.\\

\section{Results: circular polarisation}
As outlined in Sect. 2 individual pixels in an image were selected on the strength of their polarisation signal $V_{tot}$ and $LP_{tot}$. This value was calculated according to Eqs. 1 and 3 and subsequently compared to a uniform threshold that was the same for all images. The $V_{tot}$ threshold of 0.3\% corresponded to a polarisation signal of $\approx$7$\sigma$.
    
     \begin{figure}
   \centering
   \includegraphics[width=7cm]{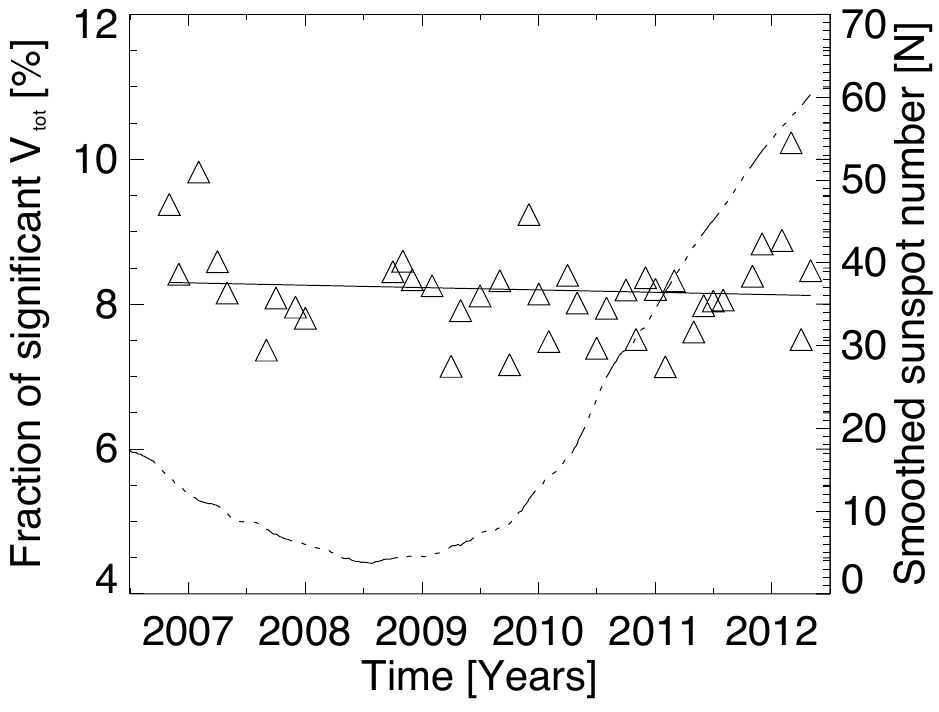}
      \caption{Fraction of pixels containing circular polarisation above a threshold of $\approx$7$\sigma$, $P_{sel} (V_{tot})$, from November 2006 until May 2012 from the original unconvolved images. The solid line is a regression with slope $-0.002 \pm 0.005$.}
         \label{VOccurrence_org}
   \end{figure}    

     \begin{figure}
   \centering
   \includegraphics[width=7cm]{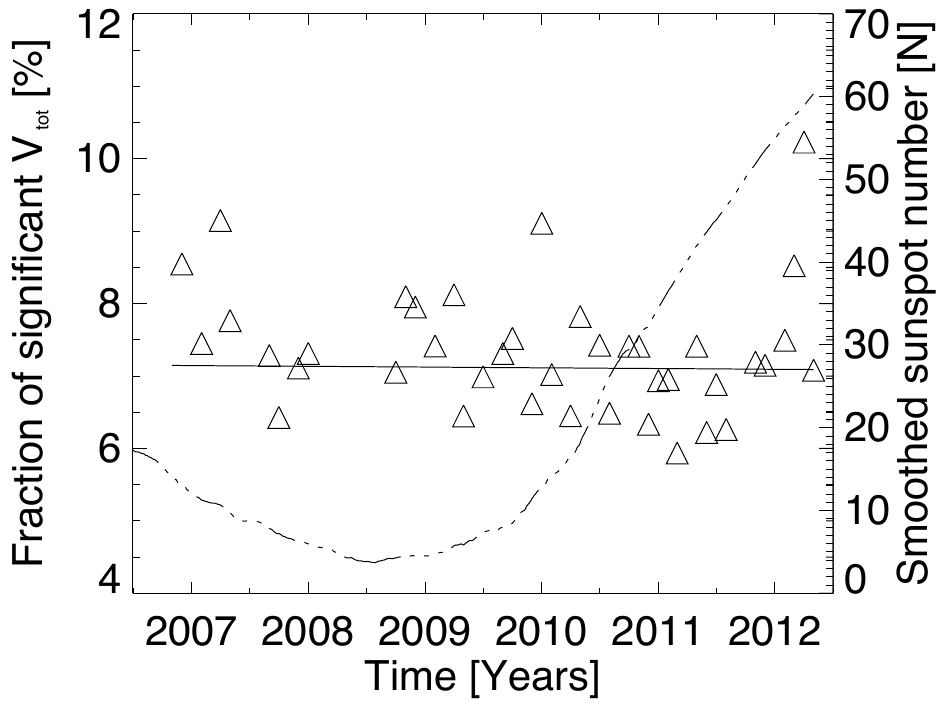}
      \caption{$P_{sel} (V_{tot})$ after each image was convolved with an appropriate artificial defocus. The solid line is a regression with slope $-0.001 \pm 0.007$.}
         \label{VOccurrence_df}
   \end{figure} 
   
First the occurrence, $P_{sel} (V_{tot})$, of the circular polarisation was investigated both, before and after the convolution of the original images with an artificial defocus. Fig. \ref{VOccurrence_org} shows the variation of $P_{sel} (V_{tot})$ from the original images i.e. before the defocus correction. The regression in this plot has a slope of $-0.002\pm 0.005$ $[\%/month]$. and indicates that there has been no significant change in the occurrence of Stokes $V$ signals. Fig. \ref{VOccurrence_df} shows the occurrence of Stokes $V$ signals at the same 0.3\% threshold after each image was convolved with an artificial defocus. As expected, the convolution has decreased the total number of selected pixels per image, but still no significant trend is discernible as the fitted regression has a slope of $-0.001\pm 0.007$ $[\%/month]$.\\  
There is a noticeable scatter between individual images in both Figs. \ref{VOccurrence_org} and \ref{VOccurrence_df}. Even with the largest images included in this study the number of large network elements varied considerably from image to image. The sudden increase in $P_{sel} (V_{tot})$ from images recorded in 2012, when compared to 2010 or 2011,  could be the result of magnetic flux from decaying active regions which are found ever closer to the disc centre as cycle 24 progresses. A quadratic fit to Figs. \ref{VOccurrence_org} and \ref{VOccurrence_df} both show a quadratic coefficient of 0.0008 $\pm$ 0.0003. However, the large variation in the number of network elements question the reliability of these fits. \\

Possible year-to-year variations in $P_{sel} (V_{tot})$ that may be due to the solar cycle were investigated further with the help of probability density functions (PDFs) for which $V_{tot} $ was converted to line-of-sight magnetic flux, Mx, using the magnetograph formula. The flux values obtained by this formula were compared to those calculated by a Milne-Eddington inversion for a single image recorded in 2007 and were found to be compatible with each other.  Distributions corresponding to each year of the investigation were produced by combining all the individual images from each year. By comparing such PDFs we can check if features with different amounts of flux all behave similarly or if e.g. larger flux features behave differently.

     \begin{figure}
   \centering
   \includegraphics[width=7cm]{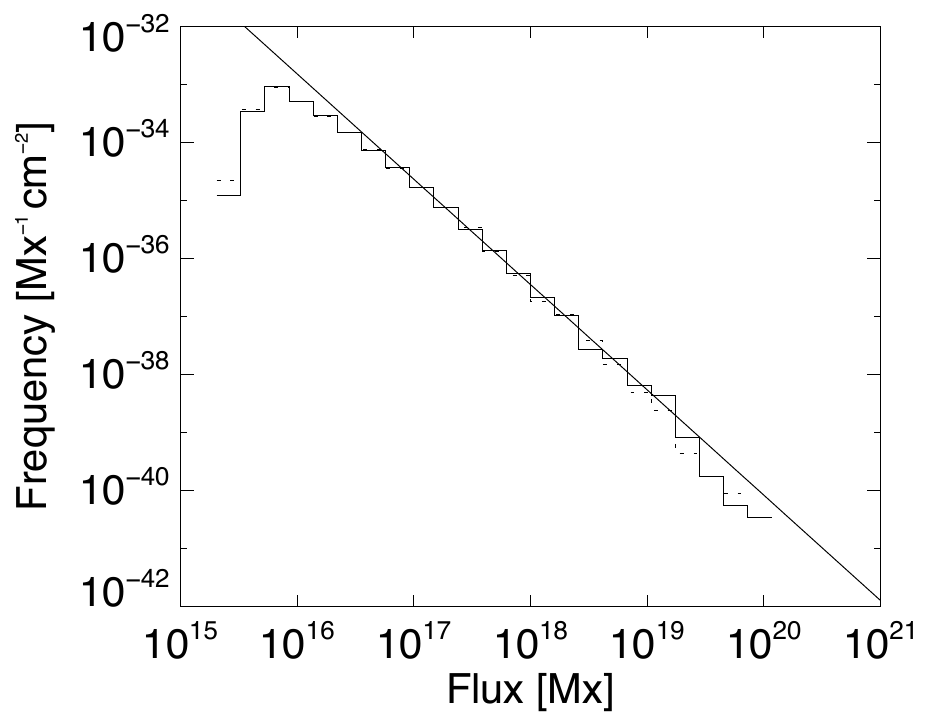}
      \caption{Two distributions of the line-of-sight magnetic flux, Mx, found on the quiet Sun. The solid line refers to a distribution from 2007, whereas the dotted line refers to 2010. The regression shows a power law fit to the 2007 distribution with spectral index $\alpha=-1.82\pm0.02$.}
         \label{logAllV}
   \end{figure}
   
     \begin{figure}
   \centering
   \includegraphics[width=7cm]{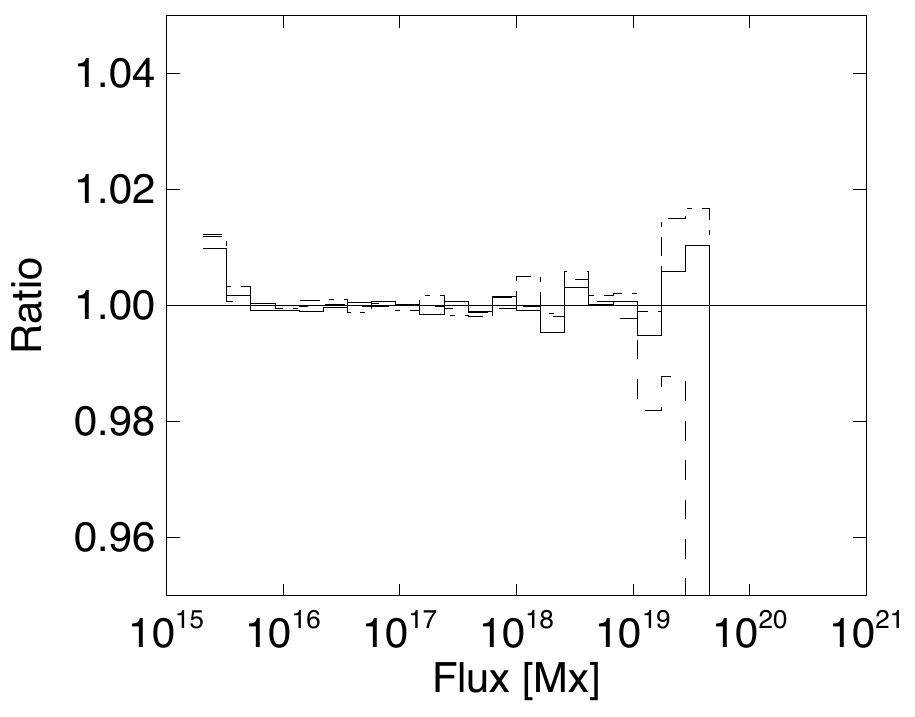}
      \caption{Ratio of three line-of-sight magnetic flux distributions, 2009: \emph{solid}, 2011: \emph{dashed}, and 2012: \emph{dot-dashed} with respect to 2007 using Eq. 4. A solid $y=1$ line has been drawn for reference.}
         \label{logAllVratio}
   \end{figure}  

Two PDFs are displayed  in Fig. \ref{logAllV} on logarithmic scales. The magnetic flux values shown on the $x$-axis correspond to flux per feature. The magnetic flux of a feature was determined by the sum of the magnetic flux found in the number of pixels that were directly connected to each other. A power law was fitted to each distribution using a least squares minimisation technique. The power law takes the form, $n(\Phi) = n_0(\Phi/\Phi_0)^\alpha$.  The PDF for 2007, solid line, has a spectral index of $\alpha= -1.82\pm0.02$, with $n_0=1.68 \times 10^{-35} Mx^{-1}cm^{-2}$ and $\Phi_0 = 1.17 \times 10^{17}$ Mx, shown by the regression in Fig. \ref{logAllV}. The PDF for 2010, dotted line, has a spectral index of $\alpha=-1.84\pm0.03$, where $n_0$ and $\Phi_0$ are the same as in 2007 up to four decimal places. Both distributions have the same binning and were fitted over the range from $n_0$ up to $8.53 \times 10^{18}$ Mx. Larger fluxes were not fitted as the number of features carrying them was very small in each year. The uncertainty for each spectral index was calculated by taking into account the uncertainty associated with every bin in the distribution over the fitted range of fluxes. By varying the value of $n_0$ for any distribution the spectral index for that distribution was seen to vary from $-1.73$ to $-1.90$. However, as long as the binning and fitted range were equal for each distribution, the variations between the yearly power law indices had a significance of only $1\sigma$ or less. The indices for the two distributions shown in Fig \ref{logAllV} agree well with the indices calculated by \citet{parnell2009} and \citet{iida2012} using SOT/NFI, despite the differing feature selection algorithms and magnetic flux calculation. The similarity between the yearly PDFs can be demonstrated further with the help of a ratio between these distributions. The ratio between two distributions was calculated using,

\begin{equation}
R = \frac{\frac{1}{ K} \sum^K_i{ f_{i}^{20xx}}}{\frac{1}{ N} \sum^N_i {f_{i}^{2007}}},
\end{equation}

where $f$ is the particular PDF of an image and both $K$ and $N$ correspond to the number of images used in the PDFs for 20xx and 2007 respectively. The distribution corresponding to 2007 was used as a 'benchmark' against which other distributions were compared, as it was composed of the largest images and hence enjoyed the best statistics. The ratios from the distributions of the years 2009, 2011 and 2012 are shown in Fig. \ref{logAllVratio} and indicate that up to fluxes of $1 \times 10^{19}$ Mx the differences between the distributions are less than 1\%, only larger fluxes show differences $> 1\%$. However, due to the small number of these large-flux features the differences bear no statistical significance. Nonetheless, differences seen for fluxes $ >1 \times 10^{19}$ Mx offer an explanation for the scatter observed in Figs. \ref{VOccurrence_org} and \ref{VOccurrence_df}, since the addition or removal of only one such feature causes a large effect in $P_{sel} (V_{tot})$ as it is composed of a large number of pixels. Ratios involving 2008 and 2010 give essentially the same result.\\

\begin{figure}
   \centering
   \includegraphics[width=7cm]{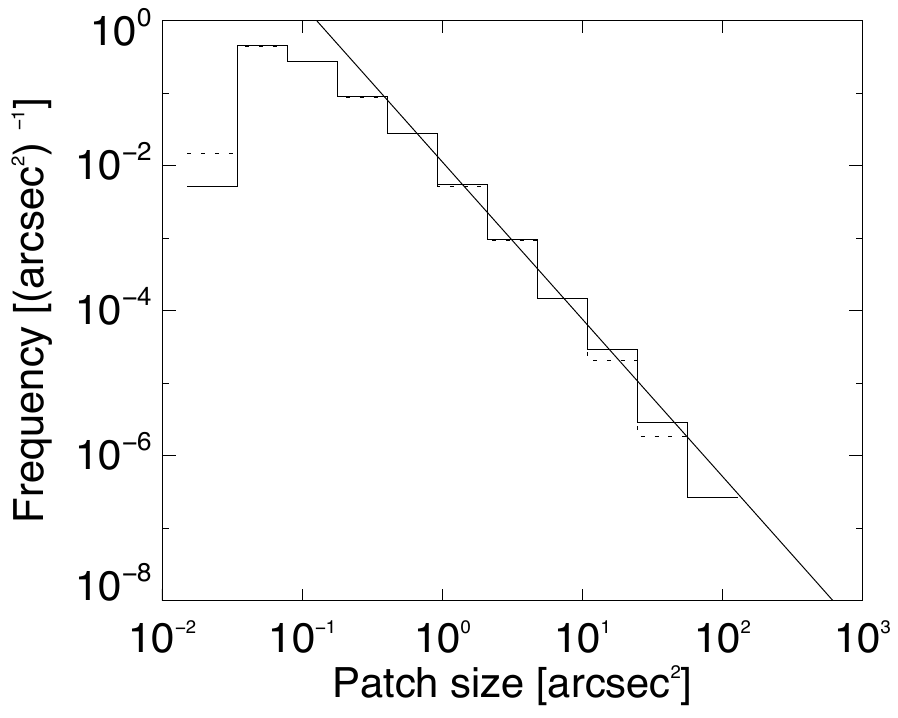}
      \caption{Two distributions of circular polarisation patch sizes. The solid line refers to a distribution from 2007, whereas the dotted line corresponds to 2010. The regression is a power law fit to the 2007 distribution with spectral index $\alpha=-2.20\pm0.05$.}
         \label{PatchV}
   \end{figure}

\begin{figure}
   \centering
   \includegraphics[width=7cm]{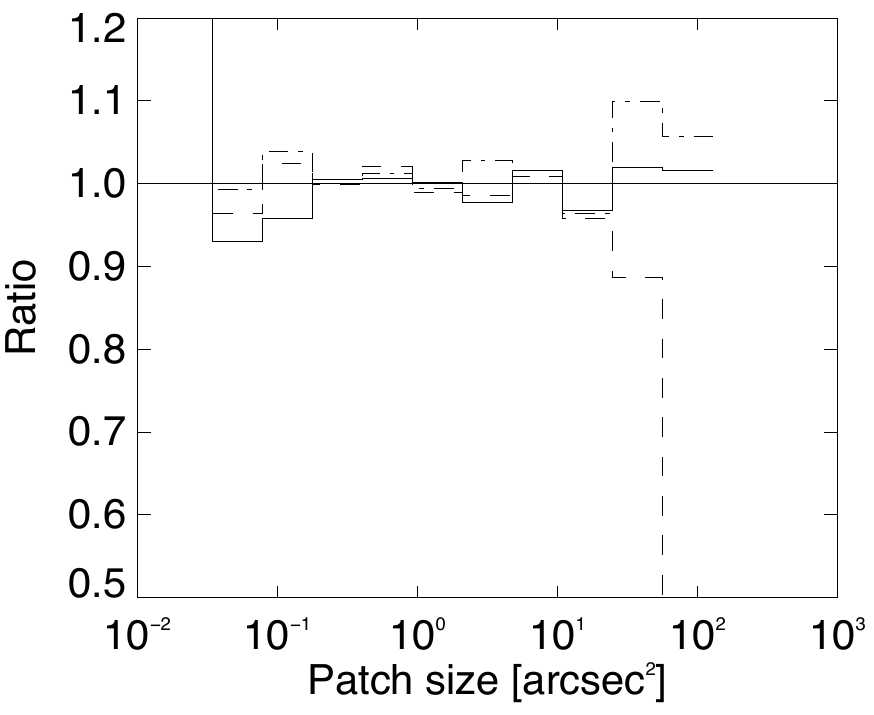}
      \caption{Ratio of three circular polarisation patch size distributions, 2009: \emph{solid}, 2011: \emph{dashed}, and 2012: \emph{dot-dashed} with respect to 2007 using Eq. 4. A solid $y=1$ line has been drawn for reference.}
         \label{PatchVratio}
   \end{figure}

This can be illustrated further by generating yearly distributions of patch sizes present on the quiet Sun. The two distributions shown in Fig. \ref{PatchV} on logarithmic scales correspond to 2007 and 2010. The distribution for 2007, solid line, has been fitted using a power law with spectral index $\alpha=-2.17\pm0.04$.  The 2010 distribution, dotted line, is also shown in Fig. \ref{PatchV} with spectral index $\alpha=-2.20\pm0.05$. The power laws in Fig. \ref{PatchV} were fitted over the range from $1.39$ arcsec$^2$ to $37.28$ arcsec$^2$.  As was the case with the magnetic flux PDFs, the yearly variations between the patch size PDFs is smaller than $1\sigma$ as long as the range of the fit and the binning is the same for every distribution. The similarity between the distributions is confirmed further by taking the ratio between them using Eq. 4. The three ratios shown in Fig. \ref{PatchVratio} used distributions from the same years as in Fig. \ref{logAllVratio}. It supports the conclusion drawn from Fig \ref{logAllVratio} that larger patches in particular those $>1$ arcsec$^2$ are mainly responsible for the scatter seen in Figs. \ref{VOccurrence_org} and \ref{VOccurrence_df}. Also, the distribution of areas drops much more rapidly than the area-integrated line-of-sight flux, illustrating that the line-of-sight flux averaged over the patch increases quickly with patch size.
    
\section{Results: linear polarisation}

The linear polarisation was investigated along the same lines as the circular polarisation in section 3. Individual pixels in an image were selected according to the strength of their average total linear polarisation signal $LP_{tot}$, calculated as described in Sect. 2. 
As was the case with the circular polarisation a constant uniform threshold was set for all images against which individual values of $LP_{tot}$ in each image were compared. The common threshold used for the linear polarisation was 0.095\%, which corresponds to a polarisation signal at $\approx$3$\sigma$. The images were also analysed using higher thresholds for $LP_{tot}$, but the results presented in this section showed no significant dependence on the choice of $LP_{tot}$. Higher thresholds severely reduced the number of selected pixels, since the number of pixels drops off very rapidly with increasing $LP_{tot}$ signal, much more rapidly than for $V_{tot}$.\\
 
\begin{figure}
   \centering
   \includegraphics[width=7cm]{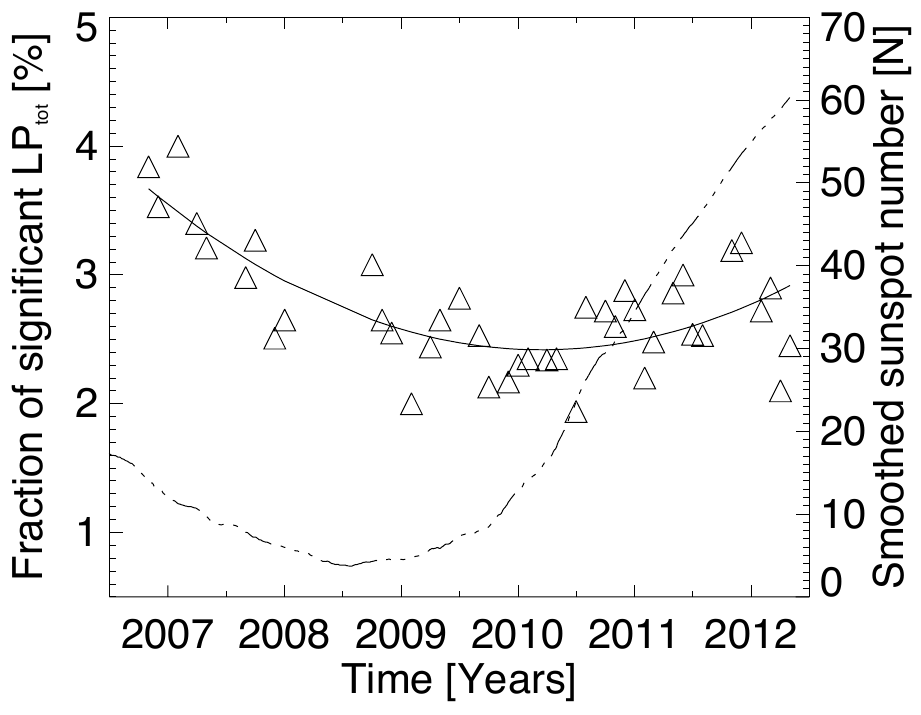}
      \caption{Fraction of pixels containing linear polarisation, $P_{sel} (LP_{tot})$, above a threshold of $\approx$3$\sigma$ from November 2006 until May 2012 from the original images. The solid line is a quadratic fit with quadratic and linear coefficients of $0.0008 \pm 0.0001$ and $-0.06 \pm 0.01$.}
         \label{LPOccurrence_org}
   \end{figure}

\begin{figure}
   \centering
   \includegraphics[width=7cm]{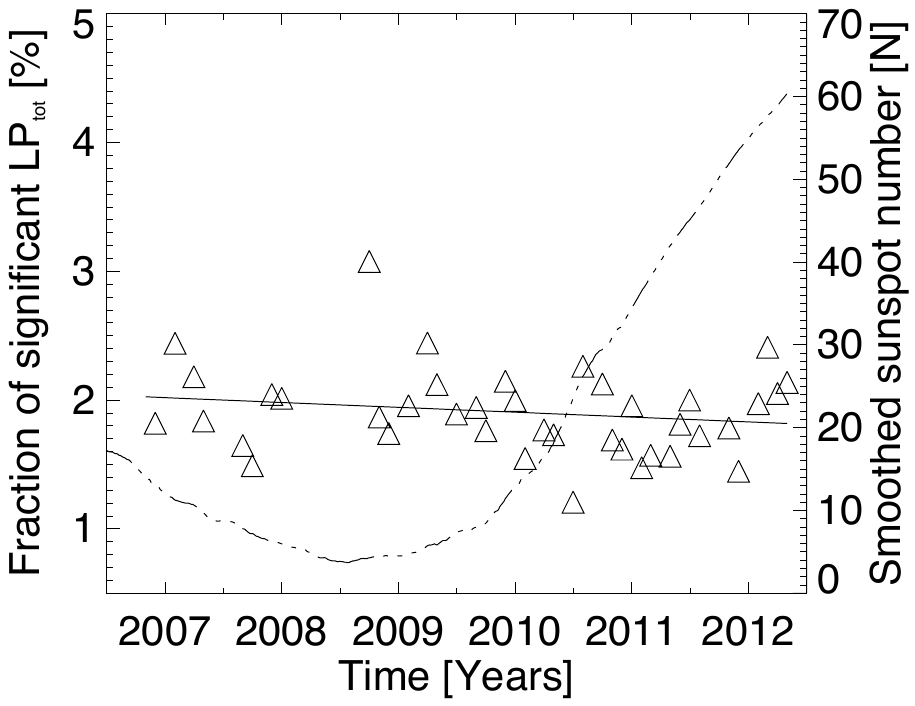}
      \caption{Fraction of pixels containing linear polarisation, $P_{sel} (LP_{tot})$, above a threshold of $\approx$3$\sigma$ after each images was convolved with an appropriate artificial defocus. The solid line is a regression with a slope of $-0.003 \pm 0.003$. The triple-dot-dash line shows the smoothed sunspot number over the same period.}
         \label{LPOccurrence_df}
   \end{figure}

Therefore results for a $LP_{tot}$ threshold other than 0.095\% are not discussed further in this section. First the variation in the occurrence, $P_{sel} (LP_{tot})$, with respect to time was investigated and is shown in Figs. \ref{LPOccurrence_org} and \ref{LPOccurrence_df}.

\begin{figure}
   \centering
   \includegraphics[width=7cm]{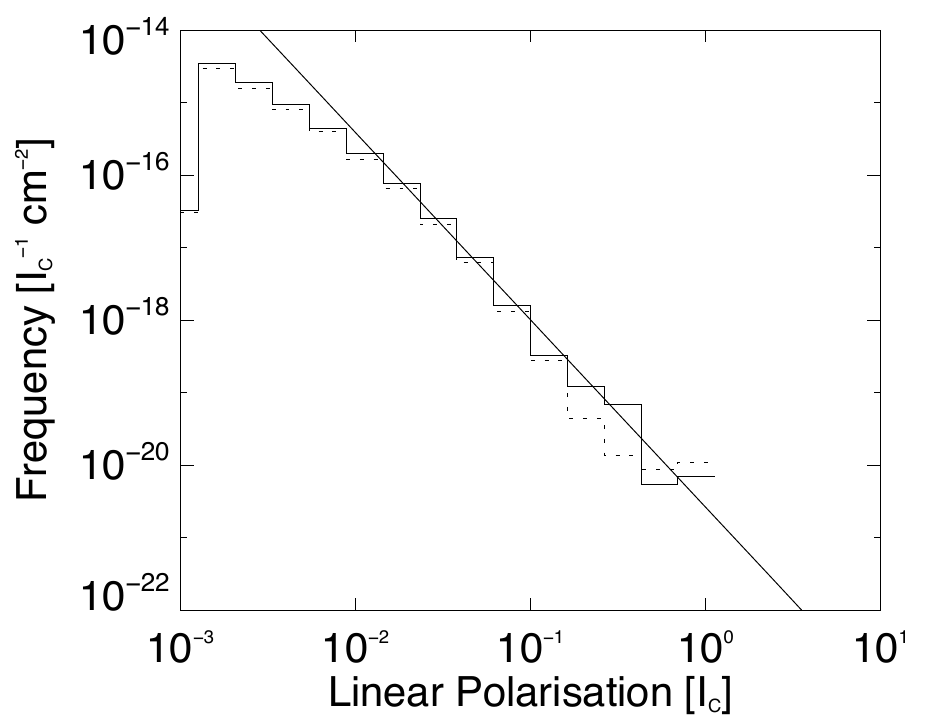}
      \caption{Distributions of the linear polarisation, $I_c$, found on the quiet Sun. The solid line refers to the distribution from 2007, whereas the dotted line refers to 2010. The regression shows a power law fit to the 2007 distribution with spectral index $\alpha=-2.60\pm0.06$.}
         \label{logAllLP}
   \end{figure}

\begin{figure}
   \centering
   \includegraphics[width=7cm]{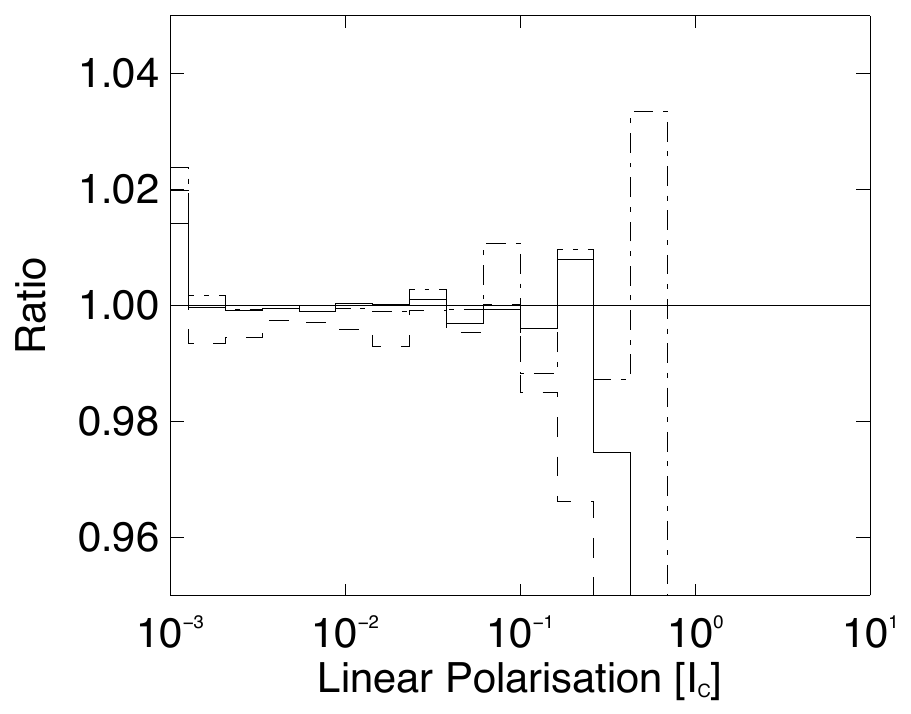}
      \caption{Ratio of three linear polarisation distributions, 2009: \emph{solid}, 2011: \emph{dashed}, and 2012: \emph{dot-dashed} with respect to 2007 using Eq. 4. A solid $y=1$ line has been drawn for reference.}
         \label{logAllLPratio}
   \end{figure}

The $P_{sel} (LP_{tot})$ in the original unconvolved images is plotted in Fig. \ref{LPOccurrence_org}. It shows that $P_{sel} (LP_{tot})$ dropped by $\approx$30\% from November 2006 until the beginning of 2010 followed by an increase until May 2012. 
The quadratic coefficient of the fit in this plot has a value of 0.0008 $\pm$ 0.0001, i.e. significant at the 8$\sigma$ level. The minimum of $P_{sel} (LP_{tot})$ took place during the beginning of 2010, which is one and half years after the official minimum (December 2008) seen in the Sunspot number. Fig. \ref{LPOccurrence_df} shows the $P_{sel} (LP_{tot})$ after each image has been convolved. The convolution has completely removed the quadratic term present in Fig. \ref{LPOccurrence_org}, demonstrating how thoroughly the contrast variation affects the observed polarisation signal. The solid regression line in Fig \ref{LPOccurrence_df} has a slope of -0.003 $\pm$ 0.003 and suggests that $P_{sel} (LP_{tot})$ has remained invariant throughout the period of investigation. However, as was the case with the circular polarisation in Figs. \ref{VOccurrence_org} and \ref{VOccurrence_df} the average $P_{sel} (LP_{tot})$ in 2010 and 2011 is lower than in 2007.\\
The decrease in $P_{sel} (LP_{tot})$, suggested in Fig. \ref{LPOccurrence_df}, was investigated further by comparing yearly PDFs of linear polarisation features. The linear polarisation of a feature was defined as the sum of the linear polarisation signal of pixels that were directly connected to each other. The distribution obtained from the data gathered in the years 2007, solid line, and 2010, dotted line, are shown in Fig. \ref{logAllLP}. Again a power law of the form $n(LP) = n_0(LP/LP_0)^\alpha$ was fitted to each distribution using a least squares minimisation technique. The power laws in Fig. \ref{logAllLP} were fitted over the range $0.018LP_{tot}$ to $0.207LP_{tot}$, where $LP_0=0.018LP_{tot}$ and $n_0=7.83 \times 10^{-17} I_c^{-1}cm^{-2}$ for the 2007 distribution and $n_0=6.51 \times 10^{-17} I_c^{-1}cm^{-2}$ for the 2010 distribution. The spectral indices are $\alpha=-2.60\pm0.06$ for 2007 and $\alpha=-2.62\pm0.09$ for 2010. The uncertainties for each spectral index took into account the uncertainties of each bin over the fitted range. Again the yearly variations of the power law indices are within $1\sigma$ as long as the binning and the fitted range is the same for each distribution. By varying $n_0$ the spectral index of a given distribution was seen to vary from $-2.34$ to $-2.87$. Even though the spectral indices of the two distributions in Fig. \ref{logAllLP} do not differ significantly from each other, a closer inspection shows that the 2010 distribution appears to be consistently lower in frequency across all bins. This can be demonstrated more clearly by taking a ratio between the yearly distributions using Eq. 4. The ratio between the 2011 and 2007 distributions plotted in Fig. \ref{logAllLPratio}, dashed line, shows that linear polarisations features with $LP_{tot}>0.0016I_c$ are consistently less frequent in 2011 than in 2007. However, at no point does this difference lie above a $1\sigma$ significance, which corresponds to $\approx4\%$ over the range $1 \times 10^{-3}$ - $1 \times 10^{-1} I_c $. The 2010 distribution shows a similar ratio. All other ratios are centered around unity as is illustrated by the two other ratios plotted in Fig. \ref{logAllLPratio}, where the solid line shows 2009 vs 2007 and the dot-dashed line corresponds to 2012 vs 2007 ratio.

\begin{figure}
   \centering
   \includegraphics[width=7cm]{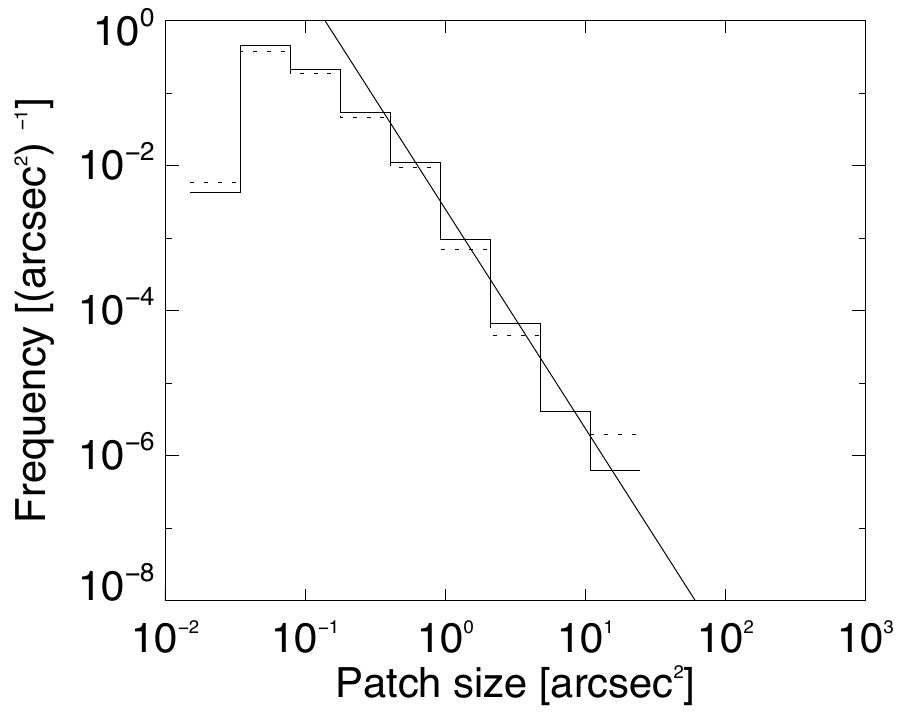}
      \caption{Two distributions of linear polarisation patch sizes. The solid line refers to a distribution from 2007, whereas the dotted line corresponds to 2010. The regression is a power law fit to the 2007 distribution with spectral index $\alpha=-3.03\pm0.08$.}
         \label{logPatch}
   \end{figure}

\begin{figure}
   \centering
   \includegraphics[width=7cm]{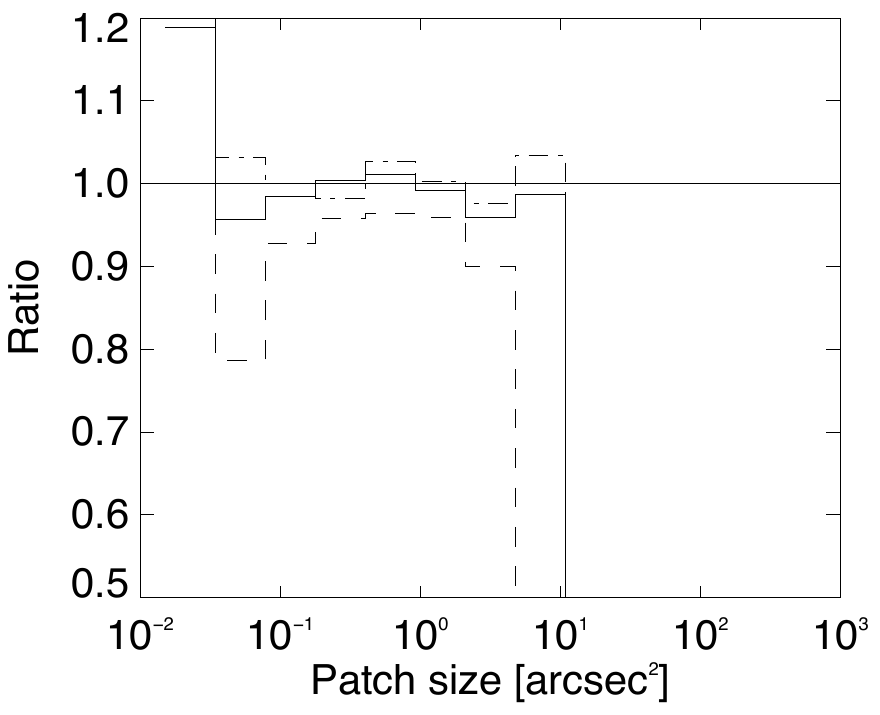}
      \caption{Ratio of three linear polarisation patch size distributions, 2009: \emph{solid}, 2011: \emph{dashed}, and 2012: \emph{dot-dashed} with respect to 2007 using Eq. 4. A solid horizontal line indicating a unit ratio has been drawn for reference.}
         \label{logPatchratio}
   \end{figure}

Finally, the yearly distributions of the linear polarisation patch sizes were examined and each distribution was fitted using a power law. The 2007 distribution, solid line, and 2010 distribution, dotted line are shown in Fig. \ref{logPatch} on logarithmic scales. The spectral indices are $\alpha=-3.03\pm0.08$ and $\alpha=-3.19\pm0.12$ for 2007 and 2010 respectively. The power laws were fitted over the range from 0.61arcsec$^2$ to 3.16 arcsec$^2$. As was the case in Fig \ref{logAllLP} linear polarisation patches in 2010 are less abundant than in 2007. This is also evident when comparing the ratio between distributions using Eq. 4 shown in Fig \ref{logPatchratio}. The three ratios correspond to the same years as for Fig \ref{logAllLPratio}. The 2011 vs 2007 ratio, dashed line in Fig \ref{logPatchratio} is below unity for all patch sizes greater than $0.052$ arcsec$^2$, the same is observed for 2010 vs 2007 ratio, not shown here. The difference has a significance of $<1\sigma$ for patches $>0.268$ arcsec$^2$ but for smaller patches a significance of $>3\sigma$ is recorded. The indices of the patch size power laws for Figs. \ref{PatchV} and \ref{logPatch} cannot be compared directly and should be interepreted with care as the power law index of such a distribution can be altered merely by a change in the polarisation thresholds for $V_{tot}$ and $LP_{tot}$. Again, the distribution of areas drops much more rapidly than the area-integrated $LP_{tot}$, indicating that $LP_{tot}$ averaged over the patch increases rapidly with patch size.

\section{Discussion and conclusion}

This investigation sought to clarify whether the number of weak magnetic elements found in the quiet Sun varies with the global solar cycle. To this end 72 Hinode SOT/SP scans of the quiet Sun  covering a period from November 2006 until May 2012 were analysed. Patches of circular and linear polarisation were analysed in a similar fashion but were treated separately. Thus, the fraction of pixels with significant polarisation signals was considered vs. time and PDFs at different epochs of the spatially averaged longitudinal or transverse field component were compared, as were PDFs of longitudinal or transverse patch sizes. \\
We discovered that the fraction of pixels with circular polarisation, $P_{sel} (V_{tot})$, above the threshold remained constant overall during the period of investigation. In particular, features with a line-of-sight magnetic  flux $<1 \times 10^{19}$ Mx showed little variation over time, indicating that magnetic flux found in these features are predominantly governed by a process that is independent of the global solar cycle. This process could take the form of an independent local dynamo action operating close to the solar surface of at least the smaller flux features \citep{ishikawa2009,pietarila2009,pietarila2010,danilovic2010,lites2011}. However, there are also other mechanisms capable of producing photospheric Zeeman signals without requiring flux emergence, such as $U$ loops \citep{pietarila2011} or the \emph{recycling} of flux from decaying active regions and network, i.e. flux produced by the global dynamo e.g. \citep{ploner2001}.
The fact that we find no variation over time for this range of magnetic flux is supported by results obtained by \citet{ito2010} and \citet{shiota2012} also using Hinode
who see the internetwork in the solar polar regions to be invariant with respect to the solar cycle. A study by \citet{kleint2010} using $C_2$ molecular-line Hanle-effect data recorded since the solar maximum of cycle 23 found some evidence to the contrary suggesting that the  weak internetwork flux may have a solar cycle dependence.\\
Possibly the good agreement of our results with those of \citet{ito2010} and \citet{shiota2012} are coincidental due to the small field of view of most of the images employed, particularly those from 2008 onwards. The majority of the images available after 2008 cover only an area corresponding to 2-3 supergranules in extend and there may be the possibility that the internetwork magnetic field could vary considerably between supergranules. However, this appears to be of only minor importance within the frame of this investigation. Since the images employed before 2008 cover a much larger area on the solar disk, those large images should provide a better representation of the \emph{average} quiet Sun. If the variation in the internetwork magnetic fields changes strongly between supergranules then the scatter in Figs. \ref{VOccurrence_org}, \ref{VOccurrence_df} and particulary \ref{LPOccurrence_org}, \ref{LPOccurrence_df} should change measurably from before 2008 to after 2008. However, as the scatter in Figs. \ref{VOccurrence_df} \& \ref{LPOccurrence_df} appears to be constant over the whole period of investigation, it can be assumed that even the small images available after 2008 are a reliable representation of the \emph{average} quiet Sun. Considering that \citet{shiota2012} not only used images covering the polar regions of the Sun but also employed images recorded at the east limb, near the solar equator, lends further support to the idea that the agreement between their and our results is not coincidental in nature.\\
Features with a magnetic flux $>1 \times 10^{19}$ Mx show more variation over time and it is the variation of these features that are mainly responsible for the scatter observed in $P_{sel} (V_{tot})$ in Figs. \ref{VOccurrence_org} and \ref{VOccurrence_df}. An analysis of the patch size of these features revealed that they occupy an area of  $> 2$ arcsec$^2$. Whilst patches $> 2$ arcsec$^2$ appear to be less abundant in 2010 and 2011 when compared to 2007, these variations are only significant at the $1\sigma$ level and therefore cannot be used as evidence for a solar cycle dependence of our observed magnetic flux. Nonetheless a significant variation in the number of these features may well be possible and has been suggested by various authors \citep{harvey1993,hagenaar2001,jin2012}. A time series extending more towards activity maximum may be required to see such variations. Furthermore, the distribution of the magnetic flux of these features can be fitted using a power law over the range from $1.17 \times 10^{17}$ Mx to $8.53 \times 10^{18}$ Mx with spectral index, in 2007, of $\alpha=-1.82\pm0.02$. This results agrees well with the results of \citet{parnell2009} and \citet{iida2012}. The index of the power law fit does not vary in time beyond a $1\sigma$ significance, at least for these quiet Sun features. The long minimum of cycle 23 and the weak cycle 24 appear to be the culprit responsible for the small variation in features with a line-of-sight flux $>1 \times 10^{19}$ Mx seen in our investigation. To what extend this unusually low activity may also have an impact on the internetwork fields, and hence on our result, is not currently clear. Note, however, that the low activity and extended length of this minimum are not unusual when considering longer term solar activity \citep[e.g.,][]{solanki2011}, so that our results may be applicable to typical minima at times of intermediate solar activity.\\
The fraction of pixels containing significant linear polarisation, $P_{sel} (LP_{tot})$, supports the observation made using circular polarisation. There is no overall variation in $P_{sel} (LP_{tot})$ over the investigated time period. This seems intuitive, since the small photospheric loops that dominate the Stokes $Q$ and $U$ signals must have corresponding foot points detectable in Stokes $V$ \citep{danilovic2010sunrise}. Therefore a potential reduction in $LP_{tot}$ should encompass a response in the circular polarisation as well. The distribution of linear polarisation features can also be fitted with a power law with spectral index, e.g. in 2007 of $\alpha=-2.60\pm0.06$. As was the case with the line of sight flux the power law index of the linear polarisation features does not vary in time beyond a significance of $1\sigma$. The distributions of linear polarisation features for the years 2010 and 2011 show that, for the range $1 \times 10^{-1}>LP_{tot}>1 \times 10^{-3}$, the frequency of features is consistently lower by 0.5\% when compared to 2007. This difference, however, was found to be insignificant for all bins even when the bins in the distributions were increased. \\
The patch size distributions of the linear polarisation also showed a drop in frequency for the years 2010 and 2011 when compared to 2007, in particular for patch sizes around $0.052$ arcsec$^2$ there was a drop of 20\% at a significance of well beyond $5\sigma$. What is surprising, is that for the line-of-sight magnetic flux the patch size distributions show no variation when compared to 2007. Also, the drop in the number of patches appears to have no significant effect on the PDF of the linear polarisation features shown in Fig. \ref{logAllLP}. \\
It needs to be said that the results obtained in this investigation strongly depend on the convolution performed on the images prior to their analysis. As was demonstrated in Figs. \ref{VOccurrence_df} and \ref{LPOccurrence_df} the effect of a variation in the continuum contrast on the observed polarisation signals is considerable. Indeed, should part of the quadratic behavior of the continuum contrast shown in Fig. \ref{spcontr} be solar in origin and not caused by the temperature variations aboard the spacecraft, the internetwork magnetic flux would inevitably show a similar behavior. A certain amount of random continuum contrast variation is almost certainly real, given that a variation in the network and internetwork should influence the number of photospheric bright points \citep{muller1984, riethmueller2010}. \\
Future work will focus on monitoring the weak quiet Sun magnetic flux over longer time periods in order to determine if the distribution of the quiet Sun magnetic flux continues to be invariant or whether the weak fields in the internetwork are affected as solar activity rises towards the next maximum. However, this task will be challenging not only by the continuing variations aboard the satellite, which in turn affect the polarimetric measurements, but also by the comparatively low frequency with which the quiet Sun magnetic fields on disk centre are recorded by SOT/SP. This problem is compounded by the fact that the solar activity is concentrated ever closer to the solar equator as cycle 24 progresses, making it difficult to ascertain the nature of the quiet Sun at the disk centre during the activity maximum. Hence, it will be interesting to investigate to what extend the Helioseismic and Magnetic Imager \citep[HMI,][]{schou2012hmi1,schou2012hmi2} aboard the Solar Dynamic Observatory (SDO) can follow internetwork magnetic flux. Its larger field-of-view is a major advantage, which might overcome some of the disadvantages caused by its lower spatial resolution. Since SDO was launched in 2010, it unfortunately missed the last activity minimum, but will be an interesting instrument to follow solar flux in the future. Nonetheless, the comparatively high spatial and spectral resolution of Hinode SOT/SP and its ability to measure all four Stokes parameters, make it a very promising instrument with which to continue investigating the internetwork magnetic field in such detail whilst covering a period of time comparable to a solar cycle.

\begin{acknowledgements}
We would like to thank Alex Feller in helping compute the \emph{ZEMAX} MTFs.
Hinode is a Japanese mission developed and launched by ISAS/JAXA, with NAOJ as domestic partner and NASA and STFC (UK) as international partners. It is operated by these agencies in co-operation with ESA and NSC (Norway).
This work has been partially supported by the WCU grant No. R31-10016 funded by the Korean Ministry of Education, Science and Technology.
D.B. acknowledges a PhD fellowship of the International Max 
Planck Research School on Physical Processes in the Solar 
System and Beyond (IMPRS).
\end{acknowledgements}

\bibliographystyle{aa}
\bibliography{TheBib_copy}{}

\begin{thebibliography}{49}
\expandafter\ifx\csname natexlab\endcsname\relax\def\natexlab#1{#1}\fi

\bibitem[{{Asensio Ramos}(2009)}]{asensio2009}
{Asensio Ramos}, A. 2009, \apj, 701, 1032

\bibitem[{{Barthol} {et~al.}(2011){Barthol}, {Gandorfer}, {Solanki},
  {Sch{\"u}ssler}, {Chares}, {Curdt}, {Deutsch}, {Feller}, {Germerott},
  {Grauf}, {Heerlein}, {Hirzberger}, {Kolleck}, {Meller}, {M{\"u}ller},
  {Riethm{\"u}ller}, {Tomasch}, {Kn{\"o}lker}, {Lites}, {Card}, {Elmore},
  {Fox}, {Lecinski}, {Nelson}, {Summers}, {Watt}, {Mart{\'{\i}}nez Pillet},
  {Bonet}, {Schmidt}, {Berkefeld}, {Title}, {Domingo}, {Gasent Blesa}, {Del
  Toro Iniesta}, {L{\'o}pez Jim{\'e}nez}, {{\'A}lvarez-Herrero},
  {Sabau-Graziati}, {Widani}, {Haberler}, {H{\"a}rtel}, {Kampf}, {Levin},
  {P{\'e}rez Grande}, {Sanz-Andr{\'e}s}, \& {Schmidt}}]{barthol2011}
{Barthol}, P., {Gandorfer}, A., {Solanki}, S.~K., {et~al.} 2011, \solphys, 268,
  1

\bibitem[{{Borrero} \& {Kobel}(2011)}]{borrero2011}
{Borrero}, J.~M. \& {Kobel}, P. 2011, \aap, 527, A29

\bibitem[{{Cattaneo}(1999)}]{cattaneo1999}
{Cattaneo}, F. 1999, \apjl, 515, L39

\bibitem[{{Danilovic} {et~al.}(2010{\natexlab{a}}){Danilovic}, {Beeck},
  {Pietarila}, {Sch{\"u}ssler}, {Solanki}, {Mart{\'{\i}}nez Pillet}, {Bonet},
  {del Toro Iniesta}, {Domingo}, {Barthol}, {Berkefeld}, {Gandorfer},
  {Kn{\"o}lker}, {Schmidt}, \& {Title}}]{danilovic2010sunrise}
{Danilovic}, S., {Beeck}, B., {Pietarila}, A., {et~al.} 2010{\natexlab{a}},
  \apjl, 723, L149

\bibitem[{{Danilovic} {et~al.}(2008){Danilovic}, {Gandorfer}, {Lagg},
  {Sch{\"u}ssler}, {Solanki}, {V{\"o}gler}, {Katsukawa}, \&
  {Tsuneta}}]{danilovic2008}
{Danilovic}, S., {Gandorfer}, A., {Lagg}, A., {et~al.} 2008, \aap, 484, L17

\bibitem[{{Danilovic} {et~al.}(2010{\natexlab{b}}){Danilovic}, {Sch{\"u}ssler},
  \& {Solanki}}]{danilovic2010}
{Danilovic}, S., {Sch{\"u}ssler}, M., \& {Solanki}, S.~K. 2010{\natexlab{b}},
  \aap, 513, A1

\bibitem[{{Fr{\"o}hlich}(2009)}]{froehlich2009}
{Fr{\"o}hlich}, C. 2009, \aap, 501, L27

\bibitem[{{Hagenaar}(2001)}]{hagenaar2001}
{Hagenaar}, H.~J. 2001, \apj, 555, 448

\bibitem[{{Harvey}(1993)}]{harvey1993}
{Harvey}, K.~L. 1993, PhD thesis, , Univ.~Utrecht, (1993)

\bibitem[{{Ichimoto} {et~al.}(2008){Ichimoto}, {Lites}, {Elmore}, {Suematsu},
  {Tsuneta}, {Katsukawa}, {Shimizu}, {Shine}, {Tarbell}, {Title}, {Kiyohara},
  {Shinoda}, {Card}, {Lecinski}, {Streander}, {Nakagiri}, {Miyashita},
  {Noguchi}, {Hoffmann}, \& {Cruz}}]{ichimoto2008sot}
{Ichimoto}, K., {Lites}, B., {Elmore}, D., {et~al.} 2008, \solphys, 249, 233

\bibitem[{{Iida} {et~al.}(2012){Iida}, {Hagenaar}, \& {Yokoyama}}]{iida2012}
{Iida}, Y., {Hagenaar}, H.~J., \& {Yokoyama}, T. 2012, \apj, 752, 149

\bibitem[{{Ishikawa} \& {Tsuneta}(2009)}]{ishikawa2009}
{Ishikawa}, R. \& {Tsuneta}, S. 2009, \aap, 495, 607

\bibitem[{{Ishikawa} {et~al.}(2008){Ishikawa}, {Tsuneta}, {Ichimoto}, {Isobe},
  {Katsukawa}, {Lites}, {Nagata}, {Shimizu}, {Shine}, {Suematsu}, {Tarbell}, \&
  {Title}}]{ishikawa2008}
{Ishikawa}, R., {Tsuneta}, S., {Ichimoto}, K., {et~al.} 2008, \aap, 481, L25

\bibitem[{{Ito} {et~al.}(2010){Ito}, {Tsuneta}, {Shiota}, {Tokumaru}, \&
  {Fujiki}}]{ito2010}
{Ito}, H., {Tsuneta}, S., {Shiota}, D., {Tokumaru}, M., \& {Fujiki}, K. 2010,
  \apj, 719, 131

\bibitem[{{Jin} \& {Wang}(2012)}]{jin2012}
{Jin}, C.~L. \& {Wang}, J.~X. 2012, \apj, 745, 39

\bibitem[{{Kleint} {et~al.}(2010){Kleint}, {Berdyugina}, {Shapiro}, \&
  {Bianda}}]{kleint2010}
{Kleint}, L., {Berdyugina}, S.~V., {Shapiro}, A.~I., \& {Bianda}, M. 2010,
  \aap, 524, A37

\bibitem[{{Kosugi} {et~al.}(2007){Kosugi}, {Matsuzaki}, {Sakao}, {Shimizu},
  {Sone}, {Tachikawa}, {Hashimoto}, {Minesugi}, {Ohnishi}, {Yamada}, {Tsuneta},
  {Hara}, {Ichimoto}, {Suematsu}, {Shimojo}, {Watanabe}, {Shimada}, {Davis},
  {Hill}, {Owens}, {Title}, {Culhane}, {Harra}, {Doschek}, \&
  {Golub}}]{kosugi2007}
{Kosugi}, T., {Matsuzaki}, K., {Sakao}, T., {et~al.} 2007, \solphys, 243, 3

\bibitem[{{Lagg} {et~al.}(2010){Lagg}, {Solanki}, {Riethm{\"u}ller},
  {Mart{\'{\i}}nez Pillet}, {Sch{\"u}ssler}, {Hirzberger}, {Feller}, {Borrero},
  {Schmidt}, {del Toro Iniesta}, {Bonet}, {Barthol}, {Berkefeld}, {Domingo},
  {Gandorfer}, {Kn{\"o}lker}, \& {Title}}]{lagg2010}
{Lagg}, A., {Solanki}, S.~K., {Riethm{\"u}ller}, T.~L., {et~al.} 2010, \apjl,
  723, L164

\bibitem[{{Lites}(2011)}]{lites2011}
{Lites}, B.~W. 2011, \apj, 737, 52

\bibitem[{{Lites} {et~al.}(2008){Lites}, {Kubo}, {Socas-Navarro}, {Berger},
  {Frank}, {Shine}, {Tarbell}, {Title}, {Ichimoto}, {Katsukawa}, {Tsuneta},
  {Suematsu}, {Shimizu}, \& {Nagata}}]{lites2008}
{Lites}, B.~W., {Kubo}, M., {Socas-Navarro}, H., {et~al.} 2008, \apj, 672, 1237

\bibitem[{{Lockwood} {et~al.}(2010){Lockwood}, {Harrison}, {Woollings}, \&
  {Solanki}}]{lockwood2010}
{Lockwood}, M., {Harrison}, R.~G., {Woollings}, T., \& {Solanki}, S.~K. 2010,
  Environmental Research Letters, 5, 024001

\bibitem[{{Mart{\'{\i}}nez Pillet} {et~al.}(2011){Mart{\'{\i}}nez Pillet}, {Del
  Toro Iniesta}, {{\'A}lvarez-Herrero}, {Domingo}, {Bonet}, {Gonz{\'a}lez
  Fern{\'a}ndez}, {L{\'o}pez Jim{\'e}nez}, {Pastor}, {Gasent Blesa}, {Mellado},
  {Piqueras}, {Aparicio}, {Balaguer}, {Ballesteros}, {Belenguer}, {Bellot
  Rubio}, {Berkefeld}, {Collados}, {Deutsch}, {Feller}, {Girela}, {Grauf},
  {Heredero}, {Herranz}, {Jer{\'o}nimo}, {Laguna}, {Meller}, {Men{\'e}ndez},
  {Morales}, {Orozco Su{\'a}rez}, {Ramos}, {Reina}, {Ramos},
  {Rodr{\'{\i}}guez}, {S{\'a}nchez}, {Uribe-Patarroyo}, {Barthol}, {Gandorfer},
  {Knoelker}, {Schmidt}, {Solanki}, \& {Vargas
  Dom{\'{\i}}nguez}}]{martinez2011}
{Mart{\'{\i}}nez Pillet}, V., {Del Toro Iniesta}, J.~C., {{\'A}lvarez-Herrero},
  A., {et~al.} 2011, \solphys, 268, 57

\bibitem[{{Muller} \& {Roudier}(1984)}]{muller1984}
{Muller}, R. \& {Roudier}, T. 1984, \solphys, 94, 33

\bibitem[{{Muller} {et~al.}(2011){Muller}, {Utz}, \& {Hanslmeier}}]{muller2011}
{Muller}, R., {Utz}, D., \& {Hanslmeier}, A. 2011, \solphys, 274, 87

\bibitem[{{Orozco Su{\'a}rez} {et~al.}(2007){Orozco Su{\'a}rez}, {Bellot
  Rubio}, {del Toro Iniesta}, {Tsuneta}, {Lites}, {Ichimoto}, {Katsukawa},
  {Nagata}, {Shimizu}, {Shine}, {Suematsu}, {Tarbell}, \& {Title}}]{orozco2007}
{Orozco Su{\'a}rez}, D., {Bellot Rubio}, L.~R., {del Toro Iniesta}, J.~C.,
  {et~al.} 2007, \apjl, 670, L61

\bibitem[{{Parnell} {et~al.}(2009){Parnell}, {DeForest}, {Hagenaar},
  {Johnston}, {Lamb}, \& {Welsch}}]{parnell2009}
{Parnell}, C.~E., {DeForest}, C.~E., {Hagenaar}, H.~J., {et~al.} 2009, \apj,
  698, 75

\bibitem[{{Petrovay} \& {Szakaly}(1993)}]{petrovay1993}
{Petrovay}, K. \& {Szakaly}, G. 1993, \aap, 274, 543

\bibitem[{{Pietarila} {et~al.}(2011){Pietarila}, {Cameron}, {Danilovic}, \&
  {Solanki}}]{pietarila2011}
{Pietarila}, A., {Cameron}, R.~H., {Danilovic}, S., \& {Solanki}, S.~K. 2011,
  \apj, 729, 136

\bibitem[{{Pietarila Graham} {et~al.}(2010){Pietarila Graham}, {Cameron}, \&
  {Sch{\"u}ssler}}]{pietarila2010}
{Pietarila Graham}, J., {Cameron}, R., \& {Sch{\"u}ssler}, M. 2010, \apj, 714,
  1606

\bibitem[{{Pietarila Graham} {et~al.}(2009){Pietarila Graham}, {Danilovic}, \&
  {Sch{\"u}ssler}}]{pietarila2009}
{Pietarila Graham}, J., {Danilovic}, S., \& {Sch{\"u}ssler}, M. 2009, \apj,
  693, 1728

\bibitem[{{Ploner} {et~al.}(2001){Ploner}, {Sch{\"u}ssler}, {Solanki}, \&
  {Gadun}}]{ploner2001}
{Ploner}, S.~R.~O., {Sch{\"u}ssler}, M., {Solanki}, S.~K., \& {Gadun}, A.~S.
  2001, in Astronomical Society of the Pacific Conference Series, Vol. 236,
  Advanced Solar Polarimetry -- Theory, Observation, and Instrumentation, ed.
  {M.~Sigwarth}, 363

\bibitem[{{Rempel} \& {Sch{\"u}ssler}(2001)}]{rempel2001}
{Rempel}, M. \& {Sch{\"u}ssler}, M. 2001, \apjl, 552, L171

\bibitem[{{Riethm{\"u}ller} {et~al.}(2010){Riethm{\"u}ller}, {Solanki},
  {Mart{\'{\i}}nez Pillet}, {Hirzberger}, {Feller}, {Bonet}, {Bello
  Gonz{\'a}lez}, {Franz}, {Sch{\"u}ssler}, {Barthol}, {Berkefeld}, {del Toro
  Iniesta}, {Domingo}, {Gandorfer}, {Kn{\"o}lker}, \&
  {Schmidt}}]{riethmueller2010}
{Riethm{\"u}ller}, T.~L., {Solanki}, S.~K., {Mart{\'{\i}}nez Pillet}, V.,
  {et~al.} 2010, \apjl, 723, L169

\bibitem[{{Schou} {et~al.}(2012{\natexlab{a}}){Schou}, {Borrero}, {Norton},
  {Tomczyk}, {Elmore}, \& {Card}}]{schou2012hmi2}
{Schou}, J., {Borrero}, J.~M., {Norton}, A.~A., {et~al.} 2012{\natexlab{a}},
  \solphys, 275, 327

\bibitem[{{Schou} {et~al.}(2012{\natexlab{b}}){Schou}, {Scherrer}, {Bush},
  {Wachter}, {Couvidat}, {Rabello-Soares}, {Bogart}, {Hoeksema}, {Liu},
  {Duvall}, {Akin}, {Allard}, {Miles}, {Rairden}, {Shine}, {Tarbell}, {Title},
  {Wolfson}, {Elmore}, {Norton}, \& {Tomczyk}}]{schou2012hmi1}
{Schou}, J., {Scherrer}, P.~H., {Bush}, R.~I., {et~al.} 2012{\natexlab{b}},
  \solphys, 275, 229

\bibitem[{{Sch{\"u}ssler} \& {V{\"o}gler}(2008)}]{schuessler2008}
{Sch{\"u}ssler}, M. \& {V{\"o}gler}, A. 2008, \aap, 481, L5

\bibitem[{{Shimizu} {et~al.}(2008){Shimizu}, {Nagata}, {Tsuneta}, {Tarbell},
  {Edwards}, {Shine}, {Hoffmann}, {Thomas}, {Sour}, {Rehse}, {Ito},
  {Kashiwagi}, {Tabata}, {Kodeki}, {Nagase}, {Matsuzaki}, {Kobayashi},
  {Ichimoto}, \& {Suematsu}}]{shimizu2008sot}
{Shimizu}, T., {Nagata}, S., {Tsuneta}, S., {et~al.} 2008, \solphys, 249, 221

\bibitem[{{Shiota} {et~al.}(2012){Shiota}, {Tsuneta}, {Shimojo}, {Sako},
  {Orozco Su{\'a}rez}, \& {Ishikawa}}]{shiota2012}
{Shiota}, D., {Tsuneta}, S., {Shimojo}, M., {et~al.} 2012, \apj, 753, 157

\bibitem[{{Solanki} {et~al.}(2010){Solanki}, {Barthol}, {Danilovic}, {Feller},
  {Gandorfer}, {Hirzberger}, {Riethm{\"u}ller}, {Sch{\"u}ssler}, {Bonet},
  {Mart{\'{\i}}nez Pillet}, {del Toro Iniesta}, {Domingo}, {Palacios},
  {Kn{\"o}lker}, {Bello Gonz{\'a}lez}, {Berkefeld}, {Franz}, {Schmidt}, \&
  {Title}}]{solanki2010}
{Solanki}, S.~K., {Barthol}, P., {Danilovic}, S., {et~al.} 2010, \apjl, 723,
  L127

\bibitem[{{Solanki} \& {Krivova}(2011)}]{solanki2011}
{Solanki}, S.~K. \& {Krivova}, N.~A. 2011, Science, 334, 916

\bibitem[{{Stenflo}(2010)}]{stenflo2010}
{Stenflo}, J.~O. 2010, \aap, 517, A37

\bibitem[{{Stenflo}(2012)}]{stenflo2012}
{Stenflo}, J.~O. 2012, \aap, 547, A93

\bibitem[{{Suematsu} {et~al.}(2008){Suematsu}, {Tsuneta}, {Ichimoto},
  {Shimizu}, {Otsubo}, {Katsukawa}, {Nakagiri}, {Noguchi}, {Tamura}, {Kato},
  {Hara}, {Kubo}, {Mikami}, {Saito}, {Matsushita}, {Kawaguchi}, {Nakaoji},
  {Nagae}, {Shimada}, {Takeyama}, \& {Yamamuro}}]{suematsu2008sot}
{Suematsu}, Y., {Tsuneta}, S., {Ichimoto}, K., {et~al.} 2008, \solphys, 249,
  197

\bibitem[{{Thornton} \& {Parnell}(2011)}]{thornton2011}
{Thornton}, L.~M. \& {Parnell}, C.~E. 2011, \solphys, 269, 13

\bibitem[{{Tsuneta} {et~al.}(2008){Tsuneta}, {Ichimoto}, {Katsukawa}, {Nagata},
  {Otsubo}, {Shimizu}, {Suematsu}, {Nakagiri}, {Noguchi}, {Tarbell}, {Title},
  {Shine}, {Rosenberg}, {Hoffmann}, {Jurcevich}, {Kushner}, {Levay}, {Lites},
  {Elmore}, {Matsushita}, {Kawaguchi}, {Saito}, {Mikami}, {Hill}, \&
  {Owens}}]{tsuneta2008sot}
{Tsuneta}, S., {Ichimoto}, K., {Katsukawa}, Y., {et~al.} 2008, \solphys, 249,
  167

\bibitem[{{Unno}(1956)}]{unno1956}
{Unno}, W. 1956, \pasj, 8, 108

\bibitem[{{Viticchi{\'e}} \& {S{\'a}nchez Almeida}(2011)}]{viticchie2010}
{Viticchi{\'e}}, B. \& {S{\'a}nchez Almeida}, J. 2011, \aap, 530, A14

\bibitem[{{V{\"o}gler} \& {Sch{\"u}ssler}(2007)}]{voegler2007}
{V{\"o}gler}, A. \& {Sch{\"u}ssler}, M. 2007, \aap, 465, L43

\end{thebibliography}

\onecolumn
\begin{longtable}{c c c c c}
\caption{\label{SOTimg} Hinode SOT/SP images}\\
\hline\hline
Image & Date & UT & X, Y\protect\footnotemark[1]& Area \protect\footnotemark[2]\\ 
\hline
\endhead
\hline
\endfoot
   1& 2006 Nov 26 & 13:14 & $-178,-56$ & $325 \times 162$ \\  
   2& 2006 Dec 19 & 11:35  & $-274,-1$ & $220 \times 162$ \\
   3& 2007 Feb 18& 03:20  & $-89,-1$ & $160 \times 162$ \\
   4& 2007 Apr 17& 15:04  & $-57,1$ & $110 \times 162$ \\
   5& 2007 May 20& 16:43 & $-56,1$ & $82 \times 81$ \\
   6& 2007 Sep 10& 07:20 & $-41,-1$ & $48 \times 162$ \\
   7& 2007 Oct 15& 17:21 & $-164,-1$ & $325 \times 162$ \\
   8& 2007 Dec 21& 00:20 & $-81,-1$ & $164 \times 162$ \\
   9& 2008 Jan 30& 23:34 & $-34,35$ & $60 \times 130$ \\
   10& 2008 Oct 22& 07:14 & $-14,24$ & $30 \times 130$\\
   11& 2008 Oct 22& 08:14 & $-14,-124$ & $30 \times 130$\\
   12& 2008 Nov 19& 03:40 & $-14,130$ & $30 \times 130$\\
   13& 2008 Nov 19& 04:10 & $-14,0$ & $30 \times 130$\\
   14& 2008 Dec 17& 04:34 & $-14,134$ & $30 \times 130$\\
   15& 2008 Dec 17& 05:43 & $-14,-35$ & $30 \times 130$\\
   16& 2009 Feb 17& 22:34 & $-14,-31$ & $30 \times 130$\\
   17& 2009 Feb 17& 23:34 & $-14,131$ & $30 \times 130$\\
   18& 2009 Apr 11& 18:05 & $-14,0$ & $30 \times 130$\\
   19& 2009 Apr 11& 16:20 & $-14,120$ & $30 \times 130$\\
   20& 2009 May 20& 18:42 & $-14,88$ & $30 \times 130$\\
   21& 2009 May 20& 19:07 & $-14,-88$ & $30 \times 130$\\
   22& 2009 Jul 28& 15:29 & $-14,-5$ & $30 \times 130$\\
   23& 2009 Jul 28& 14:07 & $-14,-204$ & $30 \times 130$\\
   24& 2009 Sep 01& 20:58 & $-14,-11$ & $30 \times 130$\\
   25& 2009 Sep 01& 21:48 & $-14,111$ & $30 \times 130$\\
   26& 2009 Oct 05& 16:04 & $-24,-29$ & $30 \times 130$\\
   27& 2009 Oct 05& 16:54 & $-24,108$ &$30 \times 130$\\
   28& 2009 Dec 21& 21:34 & $-24,125$ &$30 \times 130$\\
   29& 2009 Dec 21& 19:59 & $-24,-45$ &$30 \times 130$\\
   30& 2010 Jan 28& 23:21 & $-30,27$ &$30 \times 130$\\
   31& 2010 Jan 29& 00:34 & $-30,227$ &$30 \times 130$\\
   32& 2010 Feb 23& 19:58 & $-30,-37$ &$30 \times 130$\\
   33& 2010 Feb 23& 21:36 & $-30,123$ &$30 \times 130$\\
   34& 2010 Apr 29& 02:55 & $-30,138$ &$30 \times 130$\\
   35& 2010 Apr 29& 01:39 & $-30,-52$ &$30 \times 130$\\
   36& 2010 May 31& 01:40 & $-22,-72$ &$30 \times 130$\\
   37& 2010 May 31& 02:58 & $-22,118$ &$30 \times 130$\\
   38& 2010 Jul 03& 20:24 & $-19,112$ &$30 \times 130$\\
   39& 2010 Jul 03& 18:47 & $-19,-78$ &$30 \times 130$\\
   40& 2010 Aug 31& 14:19 & $-14,-145$ &$30 \times 130$\\
   41& 2010 Aug 31& 15:09 & $-14,45$ &$30 \times 130$\\
   42& 2010 Oct 07& 14:43 & $-25,-160$ &$30 \times 130$\\
   43& 2010 Oct 07& 16:07 & $-25,30$ &$30 \times 130$\\
   44& 2010 Nov 11& 16:31 & $-25,30$ &$30 \times 130$\\
   45& 2010 Nov 11& 15:07 & $-25,-160$ &$30 \times 130$\\
   46& 2010 Dec 12& 11:04 & $-85,-1$ & $160 \times 162$\\
   47& 2011 Jan 05& 11:53 & $-26,-55$ &$30 \times 130$\\
   48& 2011 Jan 05& 12:43 & $-26,135$ &$30 \times 130$\\
   49& 2011 Feb 02& 13:47 & $-26,-55$ &$30 \times 130$\\
   50& 2011 Feb 02& 14:37 & $-26,135$ &$30 \times 130$\\
   51& 2011 Mar 01& 14:06 & $-14,-145$ &$30 \times 130$\\
   52& 2011 Mar 01& 14:56 & $-14,45$ &$30 \times 130$\\
   53& 2011 May 03& 18:48 & $-30,-58$ &$30 \times 130$\\
   54& 2011 May 03& 20:25 & $-30,132$ &$30 \times 130$\\
   55& 2011 Jun 02& 20:55 & $-20,-82$ &$30 \times 130$\\
   56& 2011 Jun 02& 22:31 & $-20,108$ &$30 \times 130$\\
   57& 2011 Jun 30& 20:38 & $-18,4$ &$30 \times 130$\\
   58& 2011 Jun 30& 19:01 & $-18,-186$ &$30 \times 130$\\
   59& 2011 Aug 02& 22:32 & $-21,117$ &$30 \times 130$\\
   60& 2011 Aug 02& 20:56 & $-21,-73$ &$30 \times 130$\\
   61& 2011 Nov 30& 20:01 & $ -24,-159$ & $30 \times 130$\\
   62& 2011 Nov 30& 21:38 & $ -24,29$ & $30 \times 130$\\
   63& 2011 Dec 27& 23:00 & $ -24,159$ & $30 \times 130$\\
   64& 2011 Dec 27& 23:46 & $ -24,-29$ & $30 \times 130$\\
   65& 2012 Feb 02& 16:46 & $ -28,160$ & $30 \times 130$\\
   66& 2012 Feb 02& 15:22 & $ -28,-59$ & $30 \times 130$\\
   67& 2012 Feb 28& 14:57 & $ -27,-60$ & $30 \times 130$\\
   68& 2012 Feb 28& 16:21 & $ -27,130$ & $30 \times 130$\\ 
   69& 2012 Mar 31& 00:58 & $ -27,-60$ & $30 \times 130$\\
   70& 2012 Mar 31& 02:34 & $ -27,130$ & $30 \times 130$\\ 
   71& 2012 May 03& 17:28 & $ -26,-62$ & $30 \times 130$\\
   72& 2012 May 03& 19:05 & $ -26,127$ & $30 \times 130$\\ 
\footnotetext[1]{Commanded position, E,N arcseconds of disc centre. May vary from actual position.}
\footnotetext[2]{X,Y arcseconds}

\end{longtable}

\end{document}